\documentclass[fleqn,10pt]{wlscirep}
\title{Emergence of global scaling behaviour in the coupled Earth-atmosphere interaction}

\author[1,*]{Bijan Fallah}
\author[2,3,+]{Abbas Ali Saberi}
\author[1]{Sahar Sodoudi}

\affil[1]{Institute of Meteorology, Freie Universit\"{a}t of Berlin, Carl-Heinrich-Becker-Weg 6-10, 12165 Berlin, Germany}
\affil[2]{Department of Physics, College of Science, University of Tehran, Post Office Box 14395-547, Tehran, Iran}
\affil[3]{Institut f\"ur Theoretische Physik, Universit\"{a}t zu K\"oln, Z\"ulpicher Strasse 77, 50937 K\"oln, Germany}
\affil[*]{bijan.fallah@met.fu-berlin.de}
\affil[+]{ab.saberi@ut.ac.ir}

\usepackage{lineno}
\usepackage{overpic}
\usepackage{pict2e}
\usepackage{float}
\usepackage{xcolor,colortbl}
\definecolor{LightCyan}{rgb}{0.7,.7,.7}
\begin{abstract}
Scale invariance property in the global geometry of Earth may lead to a coupled interactive behaviour between various components of the climate system. One of the most interesting correlations exists between spatial statistics of the global topography and the temperature on Earth. Here we show that the power-law behaviour observed in the Earth topography via different approaches, resembles a scaling law in the global spatial distribution of independent atmospheric parameters. We report on observation of scaling behaviour of such variables characterized by distinct universal exponents. More specifically, we find that the spatial power-law behaviour in the fluctuations of the near surface temperature over the lands on Earth, shares the same universal exponent as of the global Earth topography, indicative of the global persistent role of the static geometry of Earth to control the steady state of a dynamical atmospheric field. Such a universal feature can pave the way to the theoretical 
understanding of the chaotic nature of the atmosphere coupled to the Earth's global topography. 
\end{abstract}
\begin{document}
\flushbottom
\maketitle
%
%
\thispagestyle{empty}

\section*{Introduction}
The estimation of future climate variability is of the major concerns of the recent century. Exploring the intrinsic properties of the climate system will improve the understanding of the characteristics of dynamical processes involved within the climate system so as to the downscaling procedures (\textit{e.g.}, stochastic parametrization schemes). Lower boundary conditions as a large-scale forcing may lend predictability to climate system \cite{charney1981,Palmer1994,Krishnamurthy2007,Hannachi2013}. Lower boundary conditions (\textit{e.g.}, topography, sea surface temperature, etc.) are usually prescribed in the General Circulation Models (GCMs). GCMs can realistically simulate the dominant modes of variations in the troposphere \cite{Limpasuvan1999}. Earth-atmosphere interactions between the surface and the free troposphere are mainly influenced by the characteristics of the atmospheric boundary layer. Thus, the net exchange of heat, mass and momentum are determined by the lower boundary conditions. 
Outside the boundary layer, the unresolved scales of motion may not be a big deal for the forecasting of synoptic and large-scale circulations \cite{Holton}. However, they are crucial for the mesoscale forecasting.
The Earth's surface is the main heat source and sink for the atmosphere. The inhomogeneous distribution of land properties leads to the temperature gradient within the boundary layer \cite{Whitemean2000}. The complex climate system transports heat (driven by solar insolation) from tropics to the higher latitudes. The midlatitude topographies are the main driver of Rossby waves which contribute to the heat transport in the atmosphere \cite{Hoskins1981,Held1990,Sauliere2011,Holton}. The Rossby waves control the formation of the atmospheric flows up to very remote regions and modify the wind pattern in the atmosphere \cite{Holton}. 

Previous studies revealed that, for a wide spectrum of scales, there exists a ``power law'' behaviour in the linear transects of the Earth's topography (\textit{i.e.}, $S \propto k^{-2}$, where $k$ is the wave number) \cite{VenningMeinesz1951,SAYLES1978,Pelletier1999,Gagnon2006,AliSaberi2013}. However, much less attention has been paid on the scale invariant analysis within the atmospheric system in connection with the scaling behaviour of the underlying topographic system. Power spectrum of temporal changes of atmospheric temperatures have been previously studied for different time scales \cite{Pelletier2002,Blender2003,Huybers2006,Vyushin2009}. It has been shown that the rainfall amounts do not obey an overall scaling (power-law) form within the hydrological scales \cite{Marani2005}. The power spectrum analysis is not applied to different length scales in such studies. Here, we report, for the first time, on a novel observation that the finger-print of the power-law 
relationship in the global Earth's topography 
also exists in the 
atmospheric dynamics as a result of Earth-atmosphere interaction.  

\begin{figure}[H]
\hspace{3pc}\noindent\includegraphics[width=35pc]{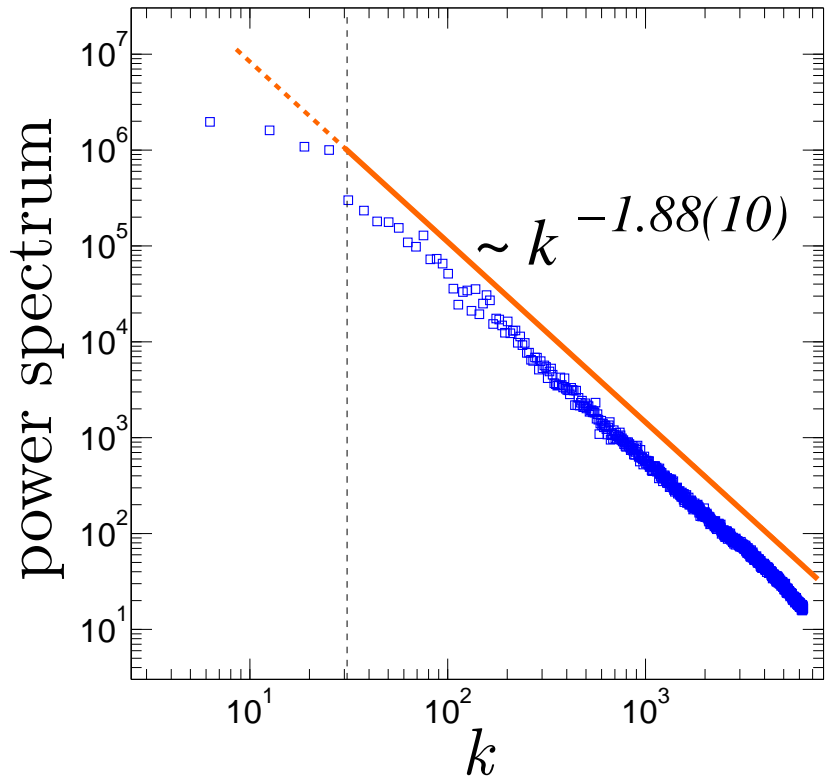}
\vspace*{0pc}
\caption{The plot of the power spectrum $S(k)$ versus the wave-number $k$ for ETOPOv2v. The solid line indicates the power-law fit to our data in the scaling region followed by the extrapolated dashed line for the large length scales.}
\label{TOPO1}
\end{figure}

\begin{figure}[H]
\hspace{2pc}\noindent\includegraphics[width=35pc]{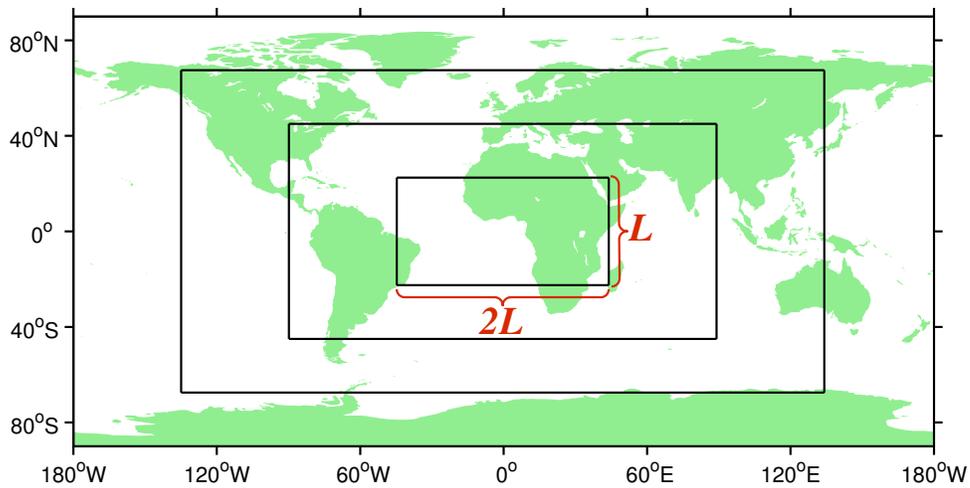}
\vspace*{0pc}
\caption{Schematic of several boxes selected for the power-law analysis. This figure is created by Mapping Toolbox of MATLAB (\url{http://www.mathworks.com/}) version R2010b.}
\label{TOPO2}
\end{figure}

\section*{Results}
\subsection*{Power Spectrum Analysis}

We reexamined power spectra as a scaling analysis of height on the ETOPO2v2 topography data. The power spectrum of linear transects of the data which contains a scaling process is given by 

\begin{equation}
S(k) \propto |k|^{-\beta}
\label{eq1}
\end{equation}  

where $\beta$ is the spectral exponent. The expected value for the exponent $\beta$ for the Earth's topography is believed to be 2 \cite{Gagnon2006,SAYLES1978}. Figure \ref{TOPO1} shows the plot of the spectral analysis of the ETOPO2v2 data set. The power spectrum is averaged over all latitudes. The power-law fit of the data presents an exponent of $\beta$= 1.88(10) (the numbers in the parenthesis show the uncertainty on the last digits), which our following cross-check shows that is more reliable than the previous estimates. In the next section, the scaling in the atmospheric surface variables is studied using a different analysis that is based on the variances of different zonal transects (different scaling) of the data. 

\subsection*{Scaling laws in the climate system}
Here, we present a new technique to capture the power-law behaviour in the climate variables (\textit{e.g.,} near surface temperatures). Prior to applying the analysis on temperature data, let us examine and validate the approach on the ETOPO2v2 orographic data.\par

\begin{figure}[H] 
\centering
\noindent\includegraphics[width=25pc]{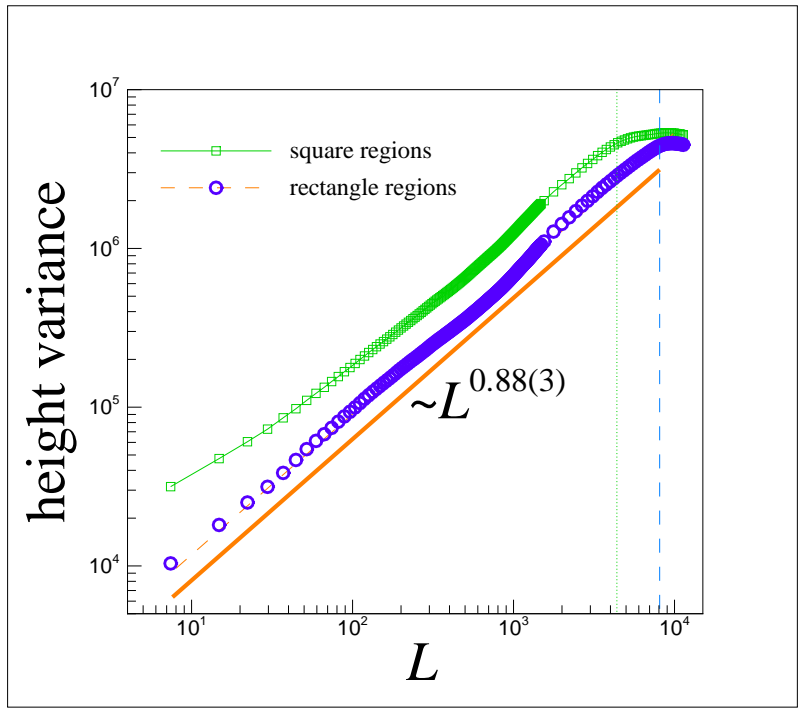}
\vspace*{3pc}
\caption{The plot of the height variances within the boxes that are randomly chosen on Earth topography (green squares for the square ($L \times L$) and blue circles for the rectangle ($2L \times L$) box sizes). The solid line shows the best power-law fit. Due to the higher number of data points within a rectangular region of size $L$, rather than a square one, we see a wider scaling interval with higher statistics for rectangular regions. That's why in our next computations, we only report on averages over rectangular regions.}
\label{TOPO3}
\end{figure}

First, different areas with equal scale ($2L\times L$) are randomly chosen on the Earth's surface (fig. \ref{TOPO2}). Then, the variance of heights is averaged over the whole ensemble of areas with scale $L$. The same approach is repeated over all the possible scales $L_{min} \leq L \leq L_{max}$, where $L_{min}$ denotes for the smallest rectangular scale containing reasonable number of data points (83 km in our case) \and $L_{max}$ equals the upper limit which covers the entire data set on Earth. Finally, the power-law fit to the graph of variances versus the scales, shows the scaling exponent in the data-set [The rectangular region of size $2L \times L$ is chosen only to cover the whole data set at the largest scale, our results, however, do not change if we choose a square region of size $L \times L$ (Fig. \ref{TOPO3})]. \par 

Figure \ref{TOPO3} shows the results of the power-law analysis for Earth topography. Vertical dashed-line indicates an upper limit for the scaling region where the box is crossing a scale before which, the boxes are randomly chosen on Earth topography and the averages are taken over a relatively large number of samples. Above this scaling region, where the box size approaches the system size, the variance is just taken over the chosen box. Therefore, we conclude that the observed deviation from the scaling law in the latter region, is due to the lack of appropriate ensemble averaging. The behaviour can indeed be obtained by extrapolating the scaling behaviour to the whole region outside the scaling region. Our discussed approach captures the scaling exponent $\sim0.88$ in the global Earth's topography which is in a good agreement with our previous estimate by the spectral analysis given by $(\beta-1)\sim0.88$.

\begin{figure}[H]
\centering
\hspace{-2pc}\noindent\includegraphics[width=26pc]{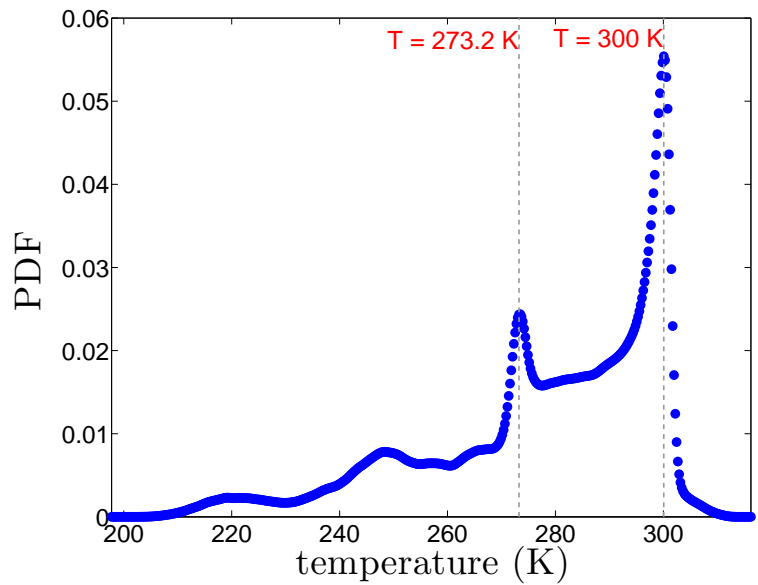}

\vspace*{0pc}
\caption{Kernel estimate of Probability Density Function of near surface temperature of the Earth from ERAInterim reanalysis.}
\label{TOPO4}
\end{figure}

\begin{figure}[H]
\centering
\noindent\includegraphics[width=25pc]{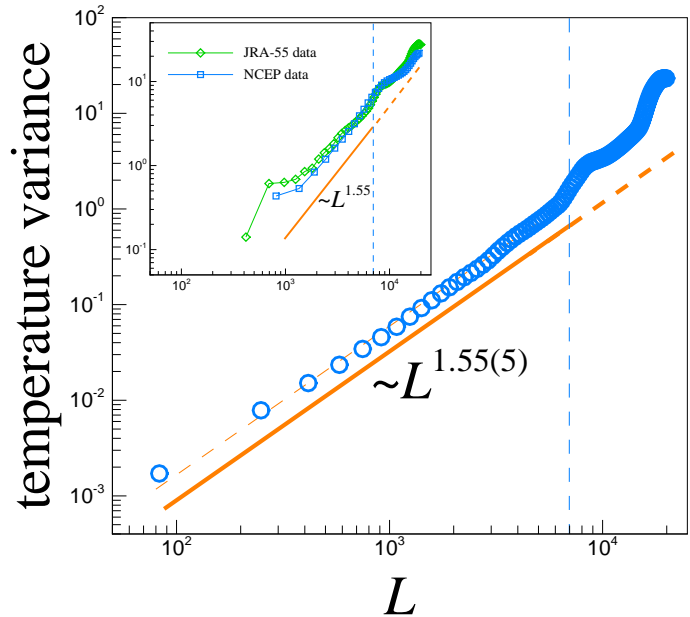}
\caption{Main: The plot of the variance of the near surface temperature fluctuations versus the length scale $L$ of the randomly chosen rectangular regions within which the temporal averages are also taken. ERAInterim data set is considered for its higher resolution and statistics. The solid line shows the best power-law fit to our data with an exponent ~1.55(5), followed by a dashed line extrapolated to larger length scales. Inset: The same analysis for other data sets JRA-55 (diamonds) and NCEP (squares) which, due to their less resolution, show a relatively poor statistics rather than the data set used in the main panel.}
\label{TOPO5}
\end{figure}

\subsubsection*{Regime behaviour in near surface air temperatures}
It has been shown [\cite{Stoddard2012} and references therein] that the topography of Earth shows a bimodal hypsography around two distinct height levels which seem to affect other geometrical observables \cite{AliSaberi2013,Saberi2015}. We find that the Earth's near surface temperature shows very similar regime behaviour. Figure \ref{TOPO4}, presents the Kernel Probability Density Function (PDF) estimate of near surface temperature from ERAInterim reanalysis. The PDF is skewed to the left and contains two distinct peaks at water's freezing point ($\sim 273.2 K$) and room temperature ($\sim 300 K$), with the latter being the global maximum. In the next step, we test the hypothesis that whether the Earth geometry is controlling the near surface atmosphere independent of horizontal scales. 

\subsubsection*{Scaling laws in the near surface temperatures}
Figure \ref{TOPO5} shows the results of the power-law analysis for near surface temperatures from ERAInterim. The universal scaling exponent for temperature variances is $\sim 1.55(5)$. As a result of sparse resolution of ERAInterim ($\sim 0.75^{\circ} \times 0.75^{\circ}$) the smallest box has a length $L_{min} = 83$ km and the ensemble size reduces after shorter iteration than in the ETOPO2v2 (dashed line in Fig. \ref{TOPO5}).  We also examined the analysis using two additional data sets, namely, i) NCEP Reanalysis 2 data \cite{NCEP} and ii) the Japanese 55-year Reanalysis (JRA-55) data \cite{JRA}. JRA-55 data has a horizontal resolution of $1.25^{\circ} \times 1.25^{\circ} $ and NCEP Reanalysis 2 data $2.5^{\circ} \times 2.5^{\circ}$ (Fig. \ref{TOPO5}).

\begin{figure}[t]
\centering\noindent\includegraphics[width=30pc]{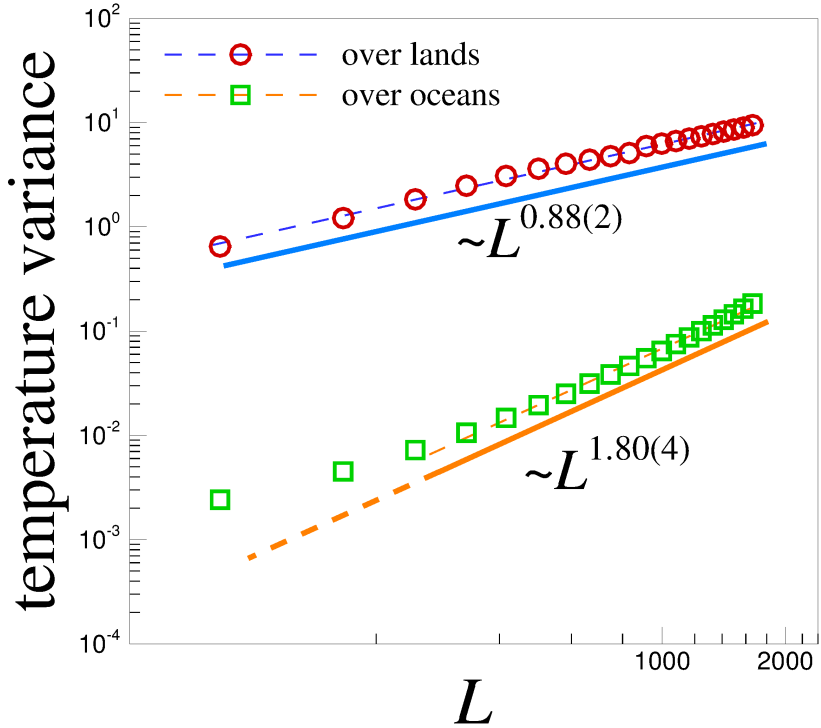}
\caption{The plot of the variance of the near surface temperature fluctuations versus the length scale L over the lands (circles) and oceans (squares). The solid lines show the best power-law fit to data in the scaling region, capturing two distinct scaling regimes over lands and oceans described by two different scaling exponents ~0.88(2) and ~1.80(4) respectively.}
\label{TOPO55_1}
\end{figure}

Let us now highlight the different role played by water on Earth to assign the scaling behaviour to the topography and the near surface temperature. One should notice that the observed scaling law in the global Earth's topography contains ‘indirect’ long-term erosive effects of water on the statistics of Earth’s bathymetry (i.e., the underwater equivalent to topography), while the global scaling behaviour in the temperature is ‘directly’ affected by the water which stabilizes and smoothens the temperature fluctuations in a wide range of length scales over the oceans. 
To further elucidate the relation between scaling exponents of topography and the near surface temperature, let us reexamine our computations for near surface temperature by separating the contributions coming from the topography over the lands and the water over the oceans (see Figure 1 in the supplementary material for further illustration). As shown in Fig. \ref{TOPO55_1}, the variance of temperature over these two regions shows a distinct difference given by the exponent $\sim 0.88(2)$ over the lands and $\sim 1.80(4)$ over the oceans. Our analysis unravels an intriguing link between the scaling behaviour observed in the near surface temperature over the lands and the scaling law for the global Earth’s topography both described by the same exponent $\sim 0.88$.

According to the fractional Brownian motion (fBm) model of self-affine interfaces, one can read the Hurst exponent H ($0\leq H<1$) from $2H^{land}\sim 0.88$ as $H^{land}\sim 0.44$ and since $H<0.5$, it shows that the temperature fluctuation field as well as the global Earth topography, display anti-persistence characterized by negative long-range correlations, i.e., a trend at $\bold{r}=(x,y)$ is not likely to be followed by a similar trend at $\bold{r}+\Delta \bold{r}$.

We also observe that the Hurst exponent for temperature fluctuations over the oceans is given by $H^{oceans} \sim 0.90$, which is close to 1, a value characteristic of a smooth fluctuating surface. Our analysis uncovers the important interplay between the Earth’s topography as a static field and the dynamical fluctuations of the near surface temperature.

\subsubsection*{Scaling laws in the other atmospheric parameters}
We also examine the existence of universal scaling in three other atmospheric parameters: mean sea level pressure, zonal surface wind and geopotential height at 500 hPa (see Figure 2 in supplementary material). Mean sea level pressure is a good proxy for weather patterns associated with different weather types and geopotential height at 500 hPa for the main tropospheric waves. Scaling exponents of several atmospheric parameters are summarized in Table \ref{tab:1}.

\begin{table}[t]
\centering

\begin{tabular}{|l|l|l|l|}
\hline
\rowcolor{LightCyan}
near surface  & near surface & mean sea level& geopotential height\\
\rowcolor{LightCyan}
temperature & zonal wind & pressure & at 500 hPa \\
\hline
1.55(5) & 0.84(10) & 1.84(20) & 1.86(10) \\
\hline
\end{tabular}
\caption{\label{tab:1} Scaling exponents of several meteorological parameters.}
\end{table}

\section*{Discussion}
Using two different approaches of scaling recognition, we have found a novel global scaling in various atmospheric parameters coupled to the statistical properties of the underlying topography of Earth. The power-law scaling behaviour in climatic variables has been studied previously, but only in the time invariant manner and for local and regional observational points. Here we employed a novel approach to extract the global scaling behaviour giving rise to some universal exponents classifying different atmospheric variables. Such scaling and possible hidden symmetries can pave the way for various exact results in such chaotic and complex systems \cite{Boffetta2008}. Existence of power-law spectrum is the base for the emergence of scale invariance property e.g., as fractal isolines (such as isothermal lines in the temperature field or coastlines in topography). Our study calls for further investigations of enrichen symmetries e.g., conformal invariance property, in the geometry of the 
fractal isothermal lines.  Possible underlying conformal invariance can then be applied to provide many interesting analytical results (for example see Ref. \cite{Boffetta2008} in which the flux of pollutant diffusing ashore from a point-like source located in the sea is analytically predicted by virtue of the discovered conformal invariance in the statistics of the iso-height lines).  \par

We propose further investigations using higher resolved atmospheric data sets in regional and global scales. The existence of power-law behaviour in atmospheric variables can, however, be applied for model validation purposes and parametrization approaches. 

\section*{Methods}
We have used the latest global atmospheric reanalysis data (6 hourly) produced by the European Centre for Medium-Range Weather Forecasts (ECMWF), the so-called ERAInterim, for our analysis \cite{ERAInterim2013,Dee2011}. The spatial resolution of the ERAInterim is $\sim 0.75^{\circ} \times 0.75^{\circ}$ ($latitude = 241\times longitude = 480$ grid points). Prior to applying the analysis, the zonal mean is excluded from the data in every time-step. The period of 1979-2013 is selected in this study. The analysis is done on the monthly means of the atmospheric near surface temperatures. By averaging the climatic features over such a long period (35 years), we investigate the mean state of the climate. We have used the 2-Minute Gridded Global Relief Data (ETOPO2v2) from U.S. Department of Commerce, National Oceanic and Atmospheric Administration, National Geophysical Data Center, (http://www.ngdc.noaa.gov/mgg/fliers/06mgg01.html) for analysis of the Earth's topography ($latitude = 5,401\times longitude = 
10,801$ grid points).


\section*{Acknowledgements}

This research was supported and funded by the University Management of Freie Universit\"{a}t Berlin. The computational resources were made available by the High Performance Computing facilities of the Freie Universit\"{a}t Berlin, Zentraleinrichtung f\"{u}r Datenverarbeitung (ZEDAT). A.A.S would like to acknowledge supports from the Alexander von Humboldt Foundation, and partial financial supports by the research council of the University of Tehran. B.F. expresses his gratitude to Dagmar M\"{u}ller for her support during production period of this manuscript.

\section*{Author contributions statement}
A.A.S. proposed the research and statistical methods. B.F. did the numerical computations and data analysis. All Authors analysed the results and drafted
the manuscript.

\section*{Additional information}
\textbf{Competing financial interests}: The authors declare no competing financial interests.




\end{document}